\documentclass[a4paper]{article}
\usepackage[utf8]{inputenc}
\usepackage[a4paper, margin=2cm]{geometry}
\usepackage{authblk}
\usepackage{url}

\title{\vspace{-1.0cm}RSEs in Research? RSEs in IT?: Finding a suitable home for RSEs}

\author[1]{Jeremy Cohen}
\author[2]{Mark Woodbridge}
\affil[1]{Department of Computing, Imperial College London}
\affil[2]{Research Computing Service, Imperial College London}

\date{}

\begin{document}

\maketitle

\abstract{
The term Research Software Engineer (RSE) was first used in the UK research community in 2012 to refer to individuals based in research environments who focus on the development of software to support and undertake research. Since then, the term has gained wide adoption and many RSE groups and teams have been set up at institutions within the UK and around the world. There is no ``manual'' for establishing an RSE group! These groups are often set up in an ad hoc manner based on what is practical within the context of each institution. Some of these groups are based in a central IT environment, others are based in research groups in academic departments. Is one option better than another? What are the pros and cons of different options? In this position paper we look at some arguments for where RSE teams fit best within academic institutions and research organisations and consider whether there is an ideal ``home'' for RSEs.
}

\section{Introduction}

The challenges faced by individuals working in a research environment but focusing on software development to support and undertake research led to the development of the Research Software Engineering (RSE) movement and the Research Software Engineer role~\cite{rse-history}. Since then, the growth of RSE has been beyond what anyone could have imagined just 8 years ago. RSE groups have been set up at institutions around the world and thriving communities have sprung up at the international, national, regional and local levels. 

Behind these communities are the RSEs themselves. They may be based within one of the many RSE groups or teams that have been set up over recent years or they may work as individuals within a research group, helping to build and maintain the software the group uses to support and undertake their research processes. Dedicated RSE groups may themselves be based either within a central institutional ICT department, or within academic departments/schools. Anecdotal evidence suggests that these groups are currently set up in the location that works best for the institution in question rather than based on any concrete analysis of the research or economic benefits of different approaches. This is unsurprising given the relatively new nature of the RSE role and the fact that it falls somewhere between a traditional research role and that of a service role such as those held by members of central IT groups. 

In this short position paper we look at the benefits and issues of RSEs and RSE groups being located in different environments within a research institution. We ask whether RSEs fit better into a purely research-based environment or a central IT environment and whether there is an ideal home for the RSE role.

\section{Research vs IT}

There are pros and cons to hosting RSEs and RSE teams within research or central IT environments. RSEs may also fit into other areas within an academic institution, for example, schools/faculties or at the institution level. However, here we consider the two most common ``homes'' for RSEs at present. 

\subsection{RSE role and group locations}

Research software is now applicable across almost all areas of research, including domains where researchers don't traditionally have experience of software development. Individual RSEs are generally embedded within research groups where they may be the only person in their group with knowledge of software development.

RSE teams tend to host a number of RSEs. They may be working to support a specific domain or they may be a more general group supporting the development of research software across multiple departments or an entire institution. Are there benefits to ``aggregating'' RSEs in groups and if so, at what level should they exist (e.g. Department, Faculty or Institution)? Having a team of generalists at the institution level can offer benefits such as flexibility, diversity of technical backgrounds and skills and the opportunity for knowledge exchange and skills development between team members. However, a downside of this approach is the ``distance'' between these RSEs and the research being undertaken. Nonetheless, as groups grow there is scope to bring in specialists who can enhance the work the group is able to do in specific areas~\cite{rse-groups}. Milewicz and Raybourn, in their case study of a large US-based software project~\cite{Milewicz18}, highlight the problems of technical language barriers between domains and challenges that include understanding and awareness. This suggests that RSEs working closer to the researchers that they are supporting is likely to be highly beneficial, in addition to having domain expertise. A team of domain specialists based within an individual department can potentially work more effectively with domain researchers on a more technical level but they may have access to a much smaller pool of work, making sustaining and growing the team more challenging.

Traditionally, many individuals doing RSE-type work within research groups have been employed as researchers, often on fixed-term contracts. There is a challenge in sustaining such roles and providing career opportunities for the RSEs undertaking them. These individuals often have specialist skills that take time to develop, so short-term, unsustainable roles present a major challenge for retention and career development.

Where RSE roles are located within a central IT environment, the ever-increasing use of software and computing infrastructure within research means that instead of being part of a self-contained RSE group, RSEs may be part of a larger IT group providing more general research support and including roles such as data scientists and High Performance Computing specialists. There is an open question as to whether RSE roles should be classed as research or professional roles -- in many cases there isn't a choice, it's down to where the roles are located or how they're funded. Where there is a choice, it's an important one.

\subsection{Benefits, challenges and tradeoffs}

We now look at some key areas, highlighting potential benefits and challenges depending on whether RSE activities are based in a central IT or research environment.\\

\noindent \textbf{Funding and career progression: }A major challenge is how RSE posts are funded. They may be funded directly by research grants as a ``directly incurred'' cost but such posts tend to be fixed-term and are unable to offer long-term job security on their own. Central funds may be more flexible and used to fund either projects onto which RSEs are recruited, or to fund RSE posts directly. Such posts are more likely to exist within central IT groups, supported by the institution as a service to researchers across a wide range of departments. The benefit of these roles is that they can, in some instances, be permanent posts that offer greater job security. As teams grow in size, there is also more scope for career advancement opportunities.\\

\noindent \textbf{Academic vs. non-academic: }Research positions generally come under an academic umbrella, potentially with different terms and salaries to non-academic roles. Research roles also generally have different performance and promotion criteria that may not be directly applicable to RSEs, e.g. paper outputs, teaching, grant income. This suggests that non-academic positions are more appropriate for RSEs. However, in the 2018 Research Software Survey, for the 6 individual countries for which results are available and the question was asked, the highest ranked reason for choosing the RSE career path was a ``Desire to work in a research environment'' or a ``Desire to advance research''~\cite{what-do-we-know, rsesurvey18}. We might ask whether a non-research role supports this. RSEs based in central teams may also be less likely to receive credit for their work, for example through co-authorship. This could, perhaps, be because these RSEs are considered to be providing a chargeable service, or because they're not seen as researchers for whom research credit is important.\\

\noindent \textbf{Skills development and knowledge exchange: }There is scope for skills development, training and knowledge exchange wherever an RSE is based. However, being part of a large group with RSEs from different research and technical backgrounds enhances this. RSEs based in academic departments are likely to have more scope to develop local research domain knowledge whereas centrally based RSEs may have more scope to learn technical skills around best practices and advanced programming techniques.\\

\noindent \textbf{Short-term vs. long-term projects: }Where small amounts of money are available for short-term software development work, hiring an RSE for a few months can be impractical. Central RSE groups are often well-placed to take on such tasks. These projects limit the opportunity for RSEs to get more involved with research teams and, potentially, contribute to papers. At the same time, short projects provide RSEs with much greater exposure to a variety of work and project types. Departmentally-based RSE groups are more likely to have RSEs funded on research grants, possibly for a number of years. This provides more scope for detailed involvement with a project but it does limit the possibility of exposure to different types of project work, reducing an RSE's ability to develop new technical skills through their day-to-day work.

\section{Conclusions}

In this article we've provided an overview of the different places that RSE roles can be located and highlighted some associated benefits and challenges. So, ``\textit{RSEs in research?}'' or ``\textit{RSEs in IT?}'', what's the answer. Well, predictably perhaps, this is not a straightforward question and the answer from our perspective is ``\textit{It depends}''. One argument is that this is, perhaps, a false dichotomy. If you want to build a large, general RSE group, then a centrally-based location, probably witin a central IT department, is probably the only option. If you want a domain-focused team of RSEs, they are likely to be best located in the department doing the work they focus on. Sometimes, regardless of the intended aim of a team, there may not be multiple options.

RSEs also need incentives to develop their skills and help others do the same through training and community engagement. Ultimately, RSEs are likely to fit best where they can be offered stable, sustainable roles with good long-term career opportunities and scope to undertake great work to support high-quality research. This is something that we hope has an increasing number of options as RSE continues to grow.

\section*{Acknowledgements}
JC acknowledges UKRI-EPSRC for support under grant EP/R025460/1.

\bibliographystyle{ieeetr}
\bibliography{rsehome-rsehpc-2020}

\end{document}